\documentclass[aps,prb,twocolumn,showpacs,preprintnumbers,amsmath,amssymb,superscriptaddress]{revtex4}

\usepackage{graphicx}
\usepackage{dcolumn}
\usepackage{bm}
\usepackage{tabularx}
\usepackage{multirow}
\usepackage{epstopdf}
\newcommand{\nc}{\newcommand}
\nc{\be}{\begin{equation}}
\nc{\ee}{\end{equation}}
\nc{\bea}{\begin{eqnarray}}
\nc{\eea}{\end{eqnarray}}
\nc{\bean}{\begin{eqnarray*}}
\nc{\eean}{\end{eqnarray*}}
\nc{\mb}{\mbox}
\nc{\rnc}{\renewcommand}
\nc{\vk}{\mb{\bf k}}
\nc{\vp}{\mb{\bf p}}
\nc{\vn}{\mb{\bf n}}
\nc{\vq}{\mb{\bf q}}
\nc{\rr}{\mb{\bf r}}
\nc{\vz}{\hat {\mb{\bf z}}}
\nc{\vj}{\mb{\boldmath$j$}}
\nc{\vg}{\mb{\boldmath$g$}}
\nc{\x}{\mb{\boldmath$x$}}
\nc{\A}{\mb{\boldmath$A$}}
\nc{\va}{\mb{\boldmath$a$}}
\nc{\vs}{\mb{\boldmath$\sigma$}}
\nc{\vpi}{\mb{\boldmath$\pi$}}
\nc{\nab}{\nabla}
\nc{\X}{\sf x}
\nc{\asp}{\hspace{10mm}}
\nc{\tg}{$t_{2g}$}
\nc{\yz}{$\{yz,zx\}$}
\nc{\STO}{SrTiO$_3$}
\nc{\KTO}{KTaO$_3$}

\begin{document}

\title{Optical conductivity of the $t_{2g}$ two-dimensional electron gas}
\author{Ming Xie}
\author{Guru Khalsa}
\author{A.H. MacDonald} 
\affiliation{Department of Physics, University of Texas at Austin, Austin TX 78712-1081, USA}

\date{\today}
\begin{abstract}
Motivated by recent interest in perovskite surfaces and heterostructures, we present an analysis of the Kubo conductivity of a two-dimensional electron gas (2DEG) formed in the $t_{2g}$ bands of 
an oxide with perovskite structure.  We find that when the electric field is polarized in the plane of the 2DEG, 
the optical conductivity is dominated by nearly independent Drude contributions from two-dimensional
subband Fermi surfaces, whereas for perpendicular-to-plane polarization it has strong 
intersubband features.  Our analysis suggests that perpendicular-to-plane optical conductivity studies
may help advance understanding of the interplay between lattice distortions and 
electron-electron interactions in complex oxide 2DEG quantum confinement physics.  
\end{abstract}


\maketitle

\section{Introduction}

Heterostructures and multilayers based on the perovskite lattice have
recently been the focus of an enormous research effort. 
The perovskite lattice supports much of the periodic table \cite{Schlom} and can be 
grown epitaxially with very high quality. \cite{Stemmer_mobility}  
Perovskite heterointerfaces often have properties that are 
drastically distinct from those of their parent bulk materials.  
The most well known example of this tendency is the polar/non-polar 
interface\cite{Hwang_LAOSTO} between the band insulators LaAlO$_3$ and SrTiO$_3$
which hosts a high mobility two-dimensional electron gas (2DEG)
formed mainly from the $t_{2g}$ orbitals of SrTiO$_3$
that can be magnetic\cite{magnetic} or superconducting\cite{super} or both.\cite{dual} 
There have by now been many studies of SrTiO$_3$ $t_{2g}$ 2DEGs formed 
at various interfaces and surfaces.\cite{OxideInterfaceReview, LTOSTO, GTOSTO, gatedSTO,Stemmer_strain}
2DEGs that are mostly similar in character but have much stronger spin-orbit (SO) coupling 
can be formed at KTaO$_3$ interfaces and surfaces.\cite{gatedKTO,Rashba}
In both SrTiO$_3$ and KTaO$_3$, it has been shown that because of 
strong and non-linear dielectric response at low temperature, the 2DEGs 
consist of a high electron
density component containing mostly electrons that are strongly confined to 
the surface or interface, and a low-density tail component consisting of weakly confined electrons
that occupy closely spaced subbands which extend $\sim$10-20 nm into the bulk of the
material.\cite{Khalsa_2DEG, Stengel_2DEG} 


Optical studies have played an important role in conventional semiconductor 2DEGs.\cite{Ando_semi, Harbeke_spec}  
Absorption of light with electric fields polarized perpendicular to the 2DEG plane has been especially valuable
because it measures intersubband optical transition energies and in this way characterizes 
2DEG quantum confinement.  No intersubband optical response is observed \cite{West} when light is 
polarized with its electric field in the 2DEG plane.  
Because optical spectra can probe intersubband transition energies,
optical characterization also has the potential to play an important role in sorting out the 
quantum confinement physics in $t_{2g}$ 2DEGs.  Experimental guidance would be 
especially valuable because of the complicating influence in the oxide case
by non-linear dielectric screening and the greater likelihood of structural distortions and defects at interfaces.
In this article we explore the optical conductivity 
of $t_{2g}$ 2DEGs theoretically, with a view toward shedding light on the information
which can be garnered from future experimental studies.\cite{Millis_optical}   
We find that the optical response of the $t_{2g}$ 2DEG is dominated by electrons within the first few layers of the surface
or interface.  When light is polarized in the 2D plane, the conductivity is dominated by 
a Drude peak to which all occupied $t_{2g}$ orbitals contribute.  The integrated strength of this 
peak provides information on the carrier density which is complementary to that available from
Hall effect measurements.  There are however weak intersubband peaks which could be very revealing 
if they could be detected.  Measurements of
the peak frequencies should be very valuable in constraining confinement models.
The corresponding peak strengths are sensitive to hybridization between 
different $t_{2g}$ orbitals, which is weak in the ideal case, and may therefore shed light on spin-orbit coupling strengths 
and on structural distortions of the pseudocubic cell near the interface.  
For light polarized perpendicular-to-plane, the optical conductivity has large intersubband features 
related to hopping amplitudes perpendicular to the interface, and to the confining potential.
In Sections ~\ref{Model} and ~\ref{linear} below we briefly discuss the model we use for a $t_{2g}$ 2DEG
and comment on the Kubo formula expressions we use for the conductivity.  Our main results are 
presented and discussed in Section ~\ref{results}.  The paper concludes in Section ~\ref{summary} with a 
brief summary and conclusions.  

\section{Model}
\label{Model}

In the perovskite ABO$_3$ unit cell, the B cation is surrounded by an octahedral oxygen cage which lifts its 
d-orbital degeneracy, pushing the $e_g$ = \{$x^2-y^2$, $3 z^2$\} levels up
relative to the $t_{2g}$ = \{$yz$, $zx$, $xy$\} levels.  
We focus on systems with conduction bands that have \tg\ character
and are well separated from oxygen p-orbital derived valance bands.  
In a cubic environment, the $yz$, $zx$, and $xy$ members of the 
$t_{2g}$ manifold are very weakly hybridized.  Under most circumstances 
atomic-like SO coupling is the dominant source of hybridization.  
When it is neglected, the three bands
therefore contribute essentially independently to physical properties of 2DEG, which 
we assume form in $x-y$ planes.   
The symmetries which lead to this circumstance can be understood by considering a two-center approximation 
description in which an $xy$ electron hops along the x-direction within a BO$_2$ plane 
from a B atom $xy$ orbital to another B atom via $\pi$ bonding to the p$_y$ orbital of the intermediate 
oxygen atom that is virtually occupied.  
We define this and other symmetry equivalent metal to metal effective hopping amplitudes as  $-t$.   
For hopping in the y-direction, the B atom $xy$ orbital hops through an oxygen p$_x$ orbital to a neighboring
B atom $xy$ orbital with the same effective hopping amplitude.
There is also a smaller but still important z-direction hopping amplitude $-t'$ for xy orbitals 
which connects one BO$_2$ layer to another that is closer to or
further from the interface.  \yz\ orbitals, on the other hand, have strong ($-t$) out-of-plane 
hopping and weak hopping ($-t'$) in one of two in-plane directions.  These orbital conserving hopping 
processes are responsible for most of the qualitative properties of \tg\ 2DEGs and are readily expressed 
mathematically by the system's tight binding model Hamiltonian.  

For a single BO$_2$ layer, the tight-binding Hamiltonian within the \tg\ subspace is,
\be
\label{TB1}
H_{SL} = \sum_{\vec{k},\gamma,\sigma} \epsilon_{\gamma} (\vec{k}) \hat{n}_{\vec{k},\gamma,\sigma}
\ee
\noindent
where $\vec{k}$ is the in-plane crystal momentum, $\gamma = \{yz,zx,xy\}$, and $\sigma$ is the spin index.  
The $\epsilon_{\gamma} (\vec{k})$ are defined by,
\bea
\label{TB2}
\nonumber
\epsilon_{yz} (\vec{k}) &=& -2 t' \cos(k_x a) - 2 t \cos(k_y a) \\ 
\epsilon_{zx} (\vec{k}) &=& -2 t \cos(k_x a)-2 t' \cos(k_y a) \\ \nonumber
\epsilon_{xy} (\vec{k}) &=& -2 t \cos(k_x a) -2 t \cos(k_y a) \\ \nonumber 
\eea 
Similarly the interlayer tunneling Hamiltonian is,
\be
H_{inter} = -\sum_{<l,l'>,\gamma,\sigma}  t_{\gamma} \; \hat{c}^{\dagger}_{\vec{k},l,\gamma,\sigma} \hat{c}_{\vec{k},l',\gamma,\sigma},
\ee
\noindent
where $<l,l'>$ are neighboring layers, $t_{\gamma}=t'$ for $\gamma=xy$, and 
$t_{\gamma}=t$ for $\gamma = \{yz,zx\}$.  
Spin-orbit coupling of the cation $d$-orbitals is mainly atomic in character.  In the calculations 
described below we have used a model\cite{Khalsa_2DEG}
in which we project atomic spin-orbit coupling onto the \tg\ space.  Note that, although 
this model contains spin-orbit coupling, it does not capture the processes which 
can lead to Rashba\cite{Rashba} spin-orbit induced momentum-splitting of the Bloch band's 
double spin degeneracy.     

Most \tg\ 2DEGs are formed near the surface of a sample sitting on a grounded substrate.
It follows that the electric field goes to zero below the 2DEG and that the electric field 
above the 2DEG is proportional to the electron density.  The electric field in the region occupied 
by the 2DEG is screened both by the 2DEG electrons and by distortions of the ionic host lattice.  
The extremely large dielectric constant of the 
host material is related to its soft anharmonic optical phonon modes, which are in turn related to the material's
nascent ferroelectricity.  We have previously described\cite{Khalsa_2DEG} a simple model which accounts for
this complex physical situation qualitatively.  We employ that model for the illustrative 
optical conductivity calculations that we report on below, forcing 
the 2DEG to lie entirely within the first 60 BO$_2$ layers.  
Because \tg\ 2DEGs can be prepared over a very wide density regime,
we have considered three representative areal densities $n$ - referred to 
below as "low" ($n = 2.3 \times 10^{13}$ cm$^{-2}$), 
"medium" ($n = 2. \times 10^{14}$ cm$^{-2}$), and "high" ($n = 5.9 \times 10^{14}$ cm$^{-2}$)
(as described in Ref. \onlinecite{Khalsa_2DEG}).
The present calculations are motivated by the expectation that quantitative 
comparisons between experiment and this simple theoretical model can be used to 
refine approximations and improve its predictive power.

\medskip
\section{Linear response theory}
\label{linear}

We consider the response of the 2DEG current to a weak external electromagnetic field. 
In the random phase approximation, the conductivity tensor is given
by the well-known Kubo formula:\cite{kubo} 
\begin{equation}
\label{conductivity}
\sigma_{\alpha\beta}(\omega) = i \hbar \sum_{m,n,\vec{k}}  \left(  \frac{f_n - f_m}{\epsilon_{m}-\epsilon_n} \right) \frac{\langle m,\vec{k}|\hat{j}_\alpha|n,\vec{k}\rangle  
\langle n,\vec{k}|\hat{j}_\beta|m,\vec{k}\rangle}{\hbar \omega  - ( \epsilon_m-\epsilon_n)  + i \eta }
\end{equation}


\noindent
where $m,n$ are band and $\alpha,\beta$ Cartesian direction indices, 
$\vec{k}$ is the 2DEG crystal momentum, and 
$\hat{j}_\alpha$ is the paramagnetic component of the current operator for which 
an explicit expression is given below. The dependence of the Fermi distribution function $f_n$ and 
the band energy $ \epsilon_n$ on $\vec{k}$ is left implicit for notational simplicity.  
The ratio of Fermi factor to energy differences should be understood as a derivative in the $m=n$ limit so
that the intraband contribution to the conductivity is 
\begin{equation}
\label{Drude}
\sigma^{IB}_{\alpha\beta}(\omega) = i \hbar \sum_{n,\vec{k}}  \left( -\frac{\partial f}{\partial \epsilon } \right) \frac{\langle n,\vec{k}|\hat{j}_\alpha|n,\vec{k}\rangle  
\langle n,\vec{k}|\hat{j}_\beta|n,\vec{k}\rangle}{\hbar \omega + i \eta }
\end{equation}
We treat $\eta=\hbar \tau^{-1}$ as a phenomenological parameter which accounts for 
the Bloch state lifetimes, assigning it a value that is independent of band index. 

In Eq.~(\ref{conductivity}) we have taken the limit $q \rightarrow 0$ because wavelengths are long
compared to lattice constants in the optical frequency regime.  The paramagnetic current operator\cite{mahan} 
is therefore given by the commutator of the Hamiltonian with the polarization operator $\hat{P}$:
\begin{equation}
\label{current}
\hat{j_\alpha} = -\frac{ie}{\hbar}[H,\hat{P}_\alpha]
\end{equation}
\noindent
In the tight binding approximation, electrons are considered to sit on lattice sites so 
position is discrete in real space.  The polarization operator therefore takes the form: $\hat{P} = \sum_i R_i \hat{n}_i$.
It follows that the in-plane current operator is given by taking the derivative of the Hamiltonian with respect to crystal momentum.  
Therefore only the $\vec{k}$ dependent part of the Hamiltonian contributes to the in-plane conductivity.  We find, from Eqs. \ref{TB1} and \ref{TB2} that the x-direction current operator is spin-independent, diagonal in layer,
and given in a $\{yz,zx,xy\}$ representation by, 

\begin{widetext}

\begin{align}
j_x = \frac{e }{\hbar}\frac{\partial H}{\partial k_x}=
 \frac{e a}{\hbar} \left(
\begin{array}{ccc}
 -2 t' \sin(k_x a) & 0 & 0 \\
 0 & -2 t \sin(k_x a) & 0 \\
 0 & 0 &  -2 t \sin(k_x a) \\
\end{array}
\right) .
\end{align} 

\end{widetext}

\noindent
Note that the in-plane current operator couples only subbands with the same orbital character and 
that its action is independent of position relative to the interface.  
In the absence of orbital hybridization (due to SO coupling in the model we considered)  
the bare Hamiltonian is also diagonal in orbital.  
It follows that in this case there are no intersubband transition contributions to the 
in-plane conductivity, either from transitions between subbands with the same orbital character or from transitions 
between subbands with different orbital character.  
When hybridization is neglected the in-plane orbital conductivity has only
a Drude response centered on $\omega \rightarrow 0$.

Because the system is finite in the z-direction, the commutator in Eq. \ref{current} is 
best evaluated in position space for this current component.  We find that 

\begin{align}
j_z= -\frac{ie}{\hbar}\sum_{\vec{k},l,\gamma} a t_{\gamma}\left( \hat{c}^\dagger_{\vec{k},l+1,\gamma}\hat{c}_{\vec{k},l,\gamma}- \hat{c}^\dagger_{\vec{k},l-1,\gamma}\hat{c}_{\vec{k},l,\gamma}\right)
\label{j3}
\end{align}

\noindent
where $t_{\gamma} = \{t,t,t' \}$ for hopping the z-direction in the $\{yz,zx,xy\}$ basis.  Because $j_z$ is off-diagonal in
layer index, optical transitions between different subbands with the same orbital character are allowed 
even in the absence of inter-orbital hybridization.  Although orbital hybridization can weakly allow 
additional optical transitions, intra-orbital contributions dominate the
perpendicular-to-plane optical response, especially so when the Fermi energy is much larger than the SO splitting.
For the calculations presented below this criterion is satisfied at medium and high densities.  
 

%
%

The real part of the longitudinal conductivity tensor satisfies
certain sum rules which are useful for verifying numerical results and 
also potentially useful in interpreting experiments.
These sum rules limit conductivity contributions from intersubband transitions.
By employing the commutation relation (\ref{current}), we obtain
the following sum rules for in-plane and perpendicular-to-plane conductivity tensors:
\begin{align}
 \int\limits_{-\infty}^{\infty}d\omega Re[\sigma_{xx}(\omega)] & = \frac{\pi e^2}{\hbar^2} \sum_{m,\vec{k}}
 \langle m \vec{k} |\frac{\partial^2 H}{\partial k_x^2}|m \vec{k} \rangle f_m \nonumber \\
 & = \pi e^2 \sum_{n} \frac{n_n}{m^*_{xx,n}}, 
 \label{sumRulexx}
\end{align}

\begin{align}
\label{sumRulezz}
\int\limits_{-\infty}^{\infty}d\omega Re[\sigma_{zz}(\omega)] = -\frac{i\pi e}{\hbar} \sum_{m,\vec{k}} 
\langle m \vec{k} |[\hat{j}_z,\hat{P}_z]|m \vec{k} \rangle f_m, 
\end{align}

\noindent
Eq. \ref{sumRulexx} is the standard result that the electronic contribution to optical conductivity integrated over all frequency is proportional to the density of electrons in that band scaled by the inverse effective mass.  The second form of the 
right hand side of Eq. \ref{sumRulexx} applies only in the low density limit in which 
the parabolic approximation for the band dispersion relations is accurate.  
Here $n_n$ corresponds to the density of the $n^{th}$ band.
In the absence of SO coupling, Eq. \ref{sumRulexx} may be simplified further to,

\begin{align}
\frac{\pi e^2}{\hbar^2} \sum_{\gamma} \frac{t_\gamma a^2}{2 } n_\gamma ,
\label{sumRuleParabolic}
\end{align}

\noindent
where $n_\gamma$ is the total density associated with electrons of orbital character $\gamma$.
The commutator in Eq. \ref{sumRulezz} is: 

\be
-\frac{i a^2}{\hbar} \sum_{\vec{k},l,\gamma} t_\gamma ( \hat{c}^{\dagger}_{\vec{k},l+1,\gamma} \hat{c}_{\vec{k},l,\gamma}  + \hat{c}^{\dagger}_{\vec{k},l-1,\gamma}  \hat{c}_{\vec{k},l,\gamma} ).
\ee

\noindent
Contributions to Eq. \ref{sumRulezz} are therefore directly proportional to the amplitude
for an electron in layer $l$ to hop to a neighboring layer, $l\pm1$.

\begin{figure}
\includegraphics[width=8.5cm,angle=0]{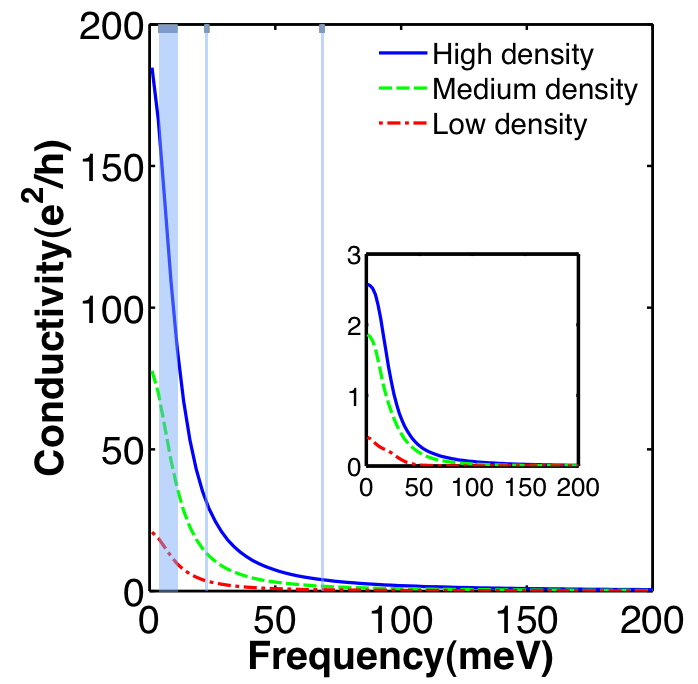} \\
\caption{(Color online) In-plane optical conductivity of a SrTiO$_3$ 2DEG
in $e^2/h$ units for light polarized in the plane of the \tg\ 2DEG at high $(5.9 \times 10^{-14}$ cm$^{-2})$, 
medium $(2 \times 10^{-14}$ cm$^{-2})$, and low $(2.3 \times 10^{-13} $ cm$^{-2})$ 
carrier densities as defined in Ref. \onlinecite{Khalsa_2DEG}.
The shaded region in the 
figure highlights the frequency region
in which the electronic conductivity is expected to be obscured by
optical phonons.  
$\eta$ has been set to $10 meV$ in order to yield $\omega \to 0$ limits 
that are similar to those observed experimentally.  The inset shows the intersubband contribution to the conductivity
which is contributed mainly by electrons in the low-density tail.
}
\label{s11}
\end{figure}

\section{Result and Discussion}   
\label{results} 
%
%

Fig. \ref{s11} shows the real part of the in-plane optical conductivity for 
$\hbar \omega$ up to $200$meV, including both the Drude conductivity and the intersubband part of the conductivity with the phenomenological scattering rate taken to be the same for both ($\eta=10$meV). 
This value of $\eta$ yields Drude peak heights in the range typical of 
recent {\em dc} resistivity measurements.\cite{resistivity} 
The Drude conductivity increases monotonically with the density of the 2DEG as 
expected from the sum rules discussed above. 
The in-plane optical conductivity is dominated by the Drude part for all densities.  
The integrated Drude weight may be used to estimate the total density, provided the effective masses in Eq. \ref{sumRuleParabolic}
can be estimated.  
From Eq. \ref{TB2} the $xy$ and $zx$ bands masses along the $x$-direction 
are larger than the $yz$ mass (by a factor $\sim$ 10).  
This suggests, that in this limit, the Drude weight will typically be dominated by the $xy$ and $zx$ pockets.
The Hall conductivity can be used to provide a complementary estimate 
of the total carrier density which weights individual orbitals differently.\cite{Us_transport}
Measurements of both quantities could be very helpful in obtaining reliable 
experimental carrier density estimates. 

As shown in the inset in Fig.\ref{s11} there is a small intersubband contribution to the conductivity
which originates from the low-density tail states.  For these states, subband separations are 
comparable to spin-orbit coupling strengths ($\Delta_{SO} = 18$ meV) allowing for considerable orbital hybridization.  
The intersubband contribution will likely be difficult to isolate experimentally because it is weak
in a relative sense.  Additionally because of the small band separations of tail states
it will be difficult to separate spectrally from the Drude peak at typical 
values of $\eta$ because it is also peaked close to $\omega=0$.
Intersubband features might be observable in systems with spin-orbit coupling 
strengths that are larger than those of 
SrTiO$_3$ ($\Delta_{SO} = 18$ meV) or in systems with substantially smaller
lifetime broadening than is currently achievable (see below). 

We remark that the utility of optical conductivity measurements as a probe of electronic properties is mitigated by the presence of 
strong optical phonon contributions.  In Fig. \ref{s11} we have shaded the frequency ranges
expected\cite{phonons} to be obscured by the three optical phonon modes which overlap with 2DEG 
energy scales.  The frequency of the low energy phonon at the $\vec{q}=0$ is strongly dependent on temperature.  To represent this we have included a shaded region spanning its temperature dependence.

\begin{figure}
\includegraphics[width=8.5cm,angle=0]{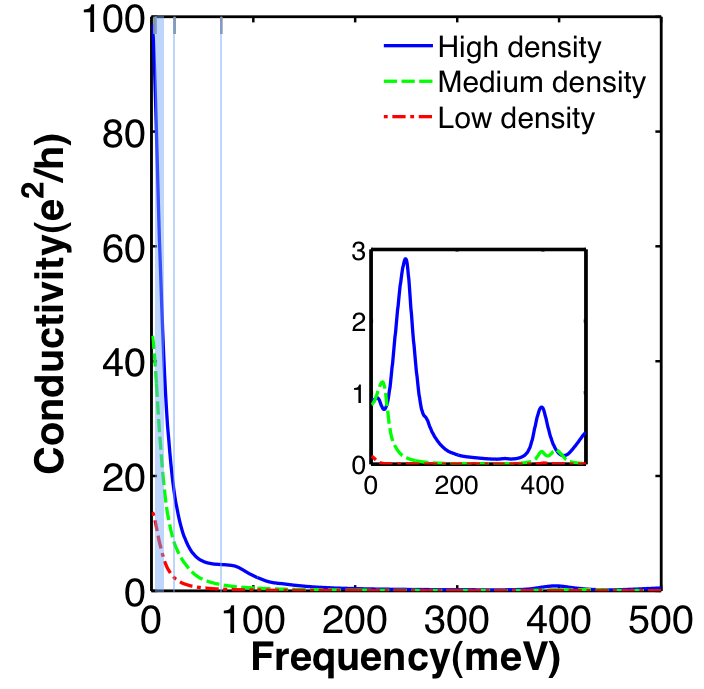} \\
\caption{(Color online) In-plane optical conductivity of a \tg\ 2DEG with strong atomic
spin-orbit coupling with strength $\Delta_{SO}=400$ meV.  
The inset plots the intersubband part only.  $\eta$ has been set to $10 meV$.}
\label{s11SO}
\end{figure}

\begin{figure*}
\includegraphics[width=17.0cm,angle=0]{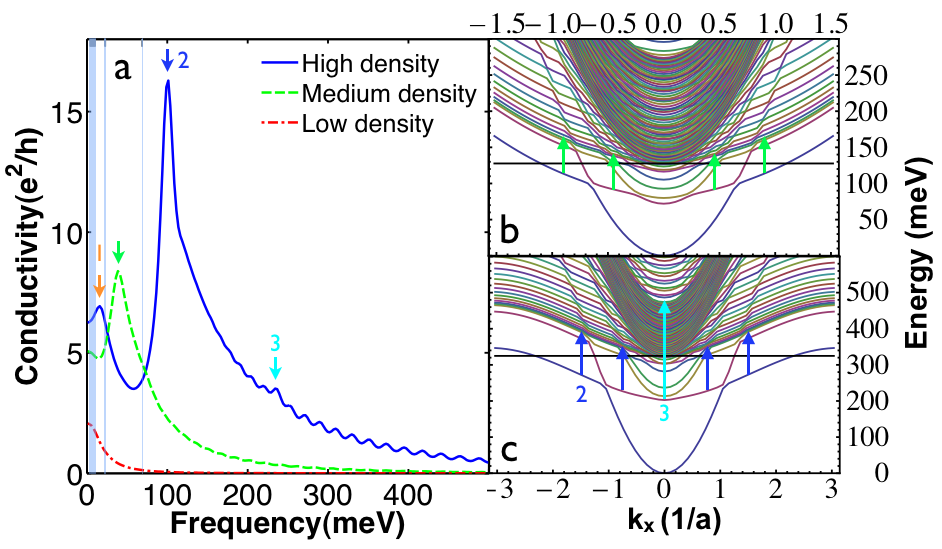} \\
\caption{(Color online)  Perpendicular-to-plane optical conductivities for low medium and high densities {\bf (a)}. 
 {\bf(b)} Self-consistent band structure for a density of $2 \times 10^{14}$cm$^{-2}$.  {\bf(c)} Self-consistent 2DEG subband structure 
 for a density of $5.9 \times 10^{14}$cm$^{-2}$.    The main optical transitions for perpendicular-to-plane polarization are
 indicated by color coordinated arrows the conductivity and bandstructure plots.  
 Feature 1 is at too low a frequency to be evident in on the energy scale of  {\bf c}.   
  The disorder-broadening $\eta$ has been set to $10$meV.}
\label{s33}
\end{figure*}

In \tg\ 2DEG systems with strong spin-orbit coupling, spectral features associated 
with transitions between strongly confined orbitals may become visible.  
To illustrate this effect, we have artificially set the atomic spin-orbit coupling strength parameter
$\Delta_{SO} \to 400$ meV in the 
SrTiO$_3$ 2DEG model and repeated the in-plane optical response calculation.
In Fig. \ref{s11SO} we see that transitions within the weakly-confined subbands
are again obscured because of the Bloch state lifetime.  Now, however, the spin-orbit coupling strength 
is strong enough to induce appreciable hybridization of the strongly-confined subbands.  
The inset of Fig. \ref{s11SO} again shows the intersubband optical response
features are most prominent at high carrier densities.    
The strongest feature is a broad peak centered at $\sim 400$ meV
associated with transitions to bulk spin-orbit split bands near the 
bottom of the conduction band.  It is present at all carrier densities, but 
stronger at higher carrier densities.  A second feature associated with the 
confinement energy scale is now allowed because of 
orbital hybridization within the $t_{2g}$ manifold.  Even in the high density case, the 400 meV SO coupling is larger than the confinement energy of the lowest subband.  This suggests that all bands have strongly hybridized \tg\ character.  Since the SO coupling is local, it does not contribute to the current operator.  Therefore, the matrix elements of the current operator in the optical conductivity still favor $xy$ and $zx$ orbitals due to their strong bonding in the $x$-direction.  The peak at $\sim 80$ meV is dominated by a peak in the $zx$ projected density of states related to the confinement energy of the most confined $\{yz,zx\}$ bands.
Because the energy scales associated with structural deformations (of the parent material or induced by the interface) 
are not expected to be this large,\cite{strain} we conclude that in the absence of large SO coupling in plane 
conductivity measurements are unlikely to provide useful information.  

Typical perpendicular-to-plane response is illustrated in Fig. \ref{s33}.  In this case 
the current operator is diagonal in orbital, but not diagonal in subband.
We therefore see a number of strong spectral features as summarized in Fig. \ref{s33}a.  
At low carrier density, the subband splittings are not much larger than the 
lifetime broadening $\eta$ and features associated with allowed transitions are 
therefore obscured in Fig. \ref{s33}a.   In practice, however, low-carrier density \tg\ 2DEGs tend to have higher mobilities,
and therefore smaller values of $\eta$ so the situation depicted in Fig. \ref{s33}a may be too 
pessimistic.  At our medium density, a peak emerges at $\sim 40$ meV 
that is associated with an optical transition from the lowest occupied heavy mass $\{yz,zx\}$ manifold 
to the many unoccupied subbands of $\{yz,zx\}$ character.  In Fig \ref{s33}b 
we show the electronic structure which yields these conductivities.  
The $\{yz,zx\}$ transitions are prominent because these bands have a large 
mass in one direction and therefore a larger density-of-states
than $xy$ bands, and also because the $z$-direction current operator is 
proportional to their larger inter-layer hopping amplitudes.    
We have highlighted the transitions responsible for the 
$\sim 40$ meV peak with green arrows in Fig. \ref{s33}b.  

For high densities, we see two features at $\sim 16$ and $\sim 100$ meV.  
We identify the higher energy features as originating from the 
$\{yz,zx\}$ transitions labelled "2" in Fig. \ref{s33}c.
(The "1" transition energy is too small to be clearly identified on the scale of
Fig. \ref{s33}c.)  The calculations also reveal many small features at higher 
energies that we associate with transitions between the most strongly 
confined $t_{2g}$ subbands and the large number of bulk-like bands due to the 60 layer simulation.\cite{Khalsa_2DEG}
These features would become sharper if we performed
our calculations with a smaller phenomenological scattering rate $\eta$.
We expect, however, that in typical systems they will yield a 
smooth tail in the optical response. 
In Fig. \ref{s33}c, highly dispersive nearly 3D $xy$ subbands can be seen with a characteristic energy width of $4 t' \sim140$ meV.  These subbands are more densely spaced at the top and bottom of this manifold due to the bonding between neighboring layers.  The non-zero SO coupling causes hybridization within the \tg\ manifold that allows \{yz,xz\} transitions to this $xy$ manifold.  The two dominant SO mediated transitions discussed above are at 100 meV and 240 meV.  The 100 meV transition gives a small contribution to the "2" peak and is therefore obscured by $\{yz,zx\}$ to $\{yz,zx\}$ transitions.  We label the "3" peak by the 240 meV transition which is not obscured by the dominant $\{yz,zx\}$ to $\{yz,zx\}$ processes.
%
%

We remark that optical spectroscopy features in conventional semiconductor 2DEGs are shifted from the
subband splittings by depolarization effects.\cite{Ando_semi,Allen}  
When electrons transition to unoccupied states, the charge distribution along the $z$ direction is altered,
which in turn alters the time-dependent electric field. Similar shifts will occur 
in oxide 2DEGs.  The peak shift is\cite{Allen} $2 e^2 n E S /\epsilon $ 
where $n$ is the density of the electrons involved in the transition, $E$ is the subband splitting and 
$S$ is an effective Coulomb interaction length\cite{Allen} for the corresponding transition.
In the conventional case, intersubband transition energies are small enough 
that $\epsilon$ can be taken to be the static dielectric constant of the host semiconductor material.
In the present case many of the transitions of interest involve strongly
confined $t_{2g}$ electrons, and have 
frequencies larger than many of the important optical phonons.  (See Fig. \ref{s33}a for example). 
In our medium density case, the transition energy falls between optical phonon modes at around
$20$ and $70$meV.  The effective dielectric constant in bulk SrTiO$_3$ in this frequency range is 
$\epsilon \sim 10$.  If we assume $10^{14}$ cm$^{-2}$ $\{yz,zx\}$ electrons are involved in the transition and 
that the Coulomb interaction length is $\sim 1$ lattice constant, we estimate that the peak shift is on the order of $190$ meV. 
This simple estimate shows that the depolarization shifts are potentially large.  
Because optical phonon oscillator strengths and frequencies are likely to be substantially
altered close to surfaces and interfaces, a fully quantitative theory of depolarization
shifts in perovskite \tg\ 2DEGs will be quite involved.

\section{Summary and Conclusions} 
\label{summary}

We report on a theoretical model study of the optical conductivity of
\tg\ 2DEGs formed at perovskite oxide surfaces and interfaces.
The detailed properties of these low-dimensional 
electron systems are difficult to predict theoretically because of the important role 
played in their properties by non-local and non-linear dielectric screening and 
by structural distortions and defects.  This study is motivated by the potential 
value of spectral and sum rule information from optical characterization for
verification and refinement of models of \tg\ 2DEG properties.

We find that the in-plane optical conductivity is very strongly dominated 
by approximately independent Drude peak contributions coming from 
all bands, irrespective of their \tg-orbital character and of the degree to 
which they are confined at the interface.  Unlike the {\em dc} conductivity, 
the Drude weight, obtained by integrating the Drude peak over frequency, 
is independent of the unknown scattering times of the various bands.  
Our calculations show that measuring the Drude weight of the
t$_{2g}$ 2DEG will provide an estimate of the total 2D carrier density that 
is complementary to the one provided by Hall effect measurements. 
The in-plane conductivity will also have features associated with 
intersubband transitions, but these will be weak unless spin-orbit interactions
hybridize \tg\ electrons with different orbital character.
In the model calculations we have performed the primary source of hybridization 
is atomic-like spin-orbit interactions which will normally have the strongest impact.
Rashba\cite{Rashba} spin-orbit interactions, which we have not specifically 
included, could provide a gate-tunable source of spin-orbit coupling which could 
enhance the value of in-plane current response measurements.

Unlike its in-plane counterpart, perpendicular-to-plane optical conductivity measurements 
should reveal a wealth of spectroscopic measurements.  Their interpretation 
will however be complicated by depolarization shifts which mean that 
spectral features cannot be immediately identified with intersubband 
transition energies.  Based on our study we conclude that 
the influence of gates, particularly the influence of back gates on ellipsometry, 
might be helpful in assigning features to particular transitions in the \tg\ 
2DEG.  When the \tg\ carrier density is reduced by a back gate\cite{Khalsa_2DEG} it has the effect of increasing the 
electric field deep in the substrate, which has a particularly strong influence on the weakly-confined 
states which are otherwise present, sharply increasing the smallest subband splittings,
decreasing the number of partially occupied subbands, and 
increasing spatial overlap between occupied and empty subbands. 
We can expect that broad tails in optical response will sharpen into 
discrete features which can be assigned on the basis of their spectral response to 
back gate voltages. 

\acknowledgments

This work was supported by NSF DMR 1122603 and by
the Robert A Welch Foundation Grant TBF1473.

%
%


%
%


\begin{thebibliography}{99}

%
%
\bibitem{Schlom} D.G. Schlom, L.Q. Chen, X. Pan, A. Schmehl, and M.A. Zurbuchen, Journal of the American Ceramic Society {\bf 91}, 2429 (2008).


\bibitem{Stemmer_mobility}
J. Son, P. Moetakef, B. Jalan, O. Bierwagen, N.J. Wright, R. Engel-Herbert, and S. Stemmer, Nat Mater {\bf 9}, 482 (2010).

%
%
\bibitem{Hwang_LAOSTO} A. Ohtomo and H.Y. Hwang, Nature {\bf 427}, 423 (2004).

%
%

\bibitem{magnetic} A. Brinkman, M. Huijben, M. van Zalk, J. Huijben, U. Zeitler, J.C. Maan, W.G. van der Wiel, G. Rjinders, D.H.A. Blank, and H. Hilgenkamp, 
Nature Materials {\bf 6}, 493 (2007).

\bibitem{super} N. Reyren, S. Thiel, A.D. Caviglia, L. Fitting Kourkoutis, G. Hammerl, C. Richter, C.W. Schneider, T. Kopp, A.S. Ruetschi, D. Jaccard, M. Gabbay, D.A. Muller, J.M Triscone, and J. Mannhart, 
Science {\bf 317}, 1196 (2007).

\bibitem{dual} L. Li, C. Richter, J. Mannhart, and R.C. Ashoori, Nature Physics {\bf 7}, 762 (2011);
D.A. Dikin, M. Mehta, C.W. Bark, C.M. Folkman, C.B. Eom, and V. Chandrasekhar, Phys. Rev. Lett. {\bf 107}, 056802 (2011);
J.A. Bert, B. Kalisky, C. Bell, M. Kim, Y. Hikita, H.Y. Hwang, and K.A. Moler, Nat. Phys. {\bf 7}, 767 (2011); 
P. Moetakef, J.R. Williams, D.G. Ouellette, A.P. Kajdos, D. Goldhaber-Gordon, S.J. Allen, and S. Stemmer, Phys. Rev. X {\bf 2}, 021014 (2012).


\bibitem{OxideInterfaceReview} J. Mannhart and D.G. Schlom, Science {\bf 327}, 1607 (2010);


J. Mannhart, D.H.A. Blank, H.Y. Hwang, A.J. Millis, and J.-M. Triscone, MRS Bulletin {\bf 33},1027 (2008).




\bibitem{LTOSTO}
A. Ohtomo, D.A. Muller, J.L. Grazul, and H.Y. Hwang, Nature {\bf 419}, 378 (2002);

%
%
\bibitem{GTOSTO}
P. Moetakef, T.A. Cain, D.G. Ouellette, J.Y. Zhang, D.O. Klenov, A. Janotti, C.G. Van de Walle, S. Rajan, S.J. Allen, and S. Stemmer, Appl. Phys. Lett. {\bf 99}, 232116 (2011); 
P. Moetakef, J.Y. Zhang, A. Kozhanov, B. Jalan, R. Seshadri, S.J. Allen, and S. Stemmer, Appl. Phys. Lett. {\bf 98}, 112110 (2011);

\bibitem{gatedSTO} 
K. Ueno, S. Nakamura, H. Shimotani, A. Ohtomo, N. Kimura, T. Nojima, H. Aoki, Y. Iwasa, and M. Kawasaki, Nature Materials {\bf 7}, 855 (2008);
Y. Lee, C. Clement, J. Hellerstedt, J. Kinney, L. Kinnischtzke, X. Leng, S.D. Snyder, and A.M. Goldman Phys. Rev. Lett. {\bf 106}, 136809 (2011);
M. Lee, J.R. Williams, S. Zhang, C.D. Frisbie, and D. Goldhaber-Gordon, Phys. Rev. Lett. {\bf 107}, 256601 (2011).

\bibitem{Stemmer_strain} B. Jalan, S.J. Allen, G.E. Beltz, P. Moetakef, and S. Stemmer, Appl. Phys. Lett. {\bf 98}, 132102 (2011).

\bibitem{gatedKTO}
K. Ueno, S. Nakamura, and H. Shimotani,  Nature Nanotechnol. {\bf 6}, 408 (2011);
P.D.C. King, R. H. He, T. Eknapakul, P. Buaphet, S.K. Mo, Y. Kaneko, S. Harashima, Y. Hikita, M.S. Bahramy, C. Bell, Z. Hussain, Y. Tokura, Z.X. Shen, H.Y. Hwang, F. Baumberger, and W. Meevasana, Phys. Rev. Lett. {\bf 108}, 117602 (2012). 

\bibitem{Rashba} 
A.D. Caviglia, M. Gabay, S. Gariglio, N. Reyren, C. Cancellieri, and J.-M. Triscone1, Phys. Rev. Lett. {\bf 104}, 126803 (2010);
A. Fete, S. Gariglio, A.D. Caviglia, J.M. Triscone, and M. Gabay, Phys. Rev. B {\bf 86}, 201105(R) (2012);
H. Nakamura, T. Koga, and T. Kimura, Phys. Rev. Lett. {\bf 108}, 206601 (2012)
Z. Zhong, A. Toth, and K. Held, Phys. Rev. B {\bf 87}, 161102(R) (2013);
G. Khalsa, B. Lee, and A.H. MacDonald, Phys. Rev. B {\bf 88}, 041302 (2013). 












\bibitem{Khalsa_2DEG} G. Khalsa and A.H. MacDonald, Phys. Rev. B, {\bf 86} 125121 (2012).

\bibitem{Stengel_2DEG} M. Stengel, Phys. Rev. Lett. {\bf 106}, 136803 (2011).


\bibitem{Ando_semi} T. Ando, A.B. Fowler, and F. Stern, Rev. Mod. Phys. {\bf 54}, 437 (1982).

\bibitem{Harbeke_spec} G. Harbeke, Physica Scripta {\bf T29}, 135 (1989).

\bibitem{West} L.C. West and S.J. Eglash, Appl. Phys. Lett. {\bf 46}, 1156 (1985).

\bibitem{Millis_optical} A complementary theoretical paper with a similar motivation appeared as we were 
preparing this paper for publication, S.Y. Park, and A.J. Millis, Phys. Rev. B {\bf 87}, 205145 (2013).
These authors made an effort to connect more directly with ellipsiometry, and in particular 
estimated the interaction induced shifts in intersubband transition energies.  Our paper 
focuses more on features associated with the weak coupling between $t_{2g}$ bands 
which can make optical characterization more useful, and accounts for their 
finite heavy-masses which often play an essential role.

\bibitem{kubo} R. Kubo, J. Phys. Soc. Jpn. {\bf 12}, 570 (1957).
\bibitem{mahan} G. Mahan, Many particle physics, Springer 2010.

\bibitem{resistivity} Z.Q. Liu, C.J. Li, W.M. Lu, X.H. Huang, Z. Huang, S.W. Zeng, X.P. Qiu, L.S. Huang, A. Annadi, J.S. Chen, J.M.D. Coey, T. Venkatesan, and Ariando, Phy. Rev. X {\bf 3}, 021010 (2013).

\bibitem{Us_transport} Ming Xie, Guru Khalsa, and A.H. MacDonald, 
to be submitted.

\bibitem{phonons} 
J.L.M. van Mechelen, D. van der Marel, C. Grimaldi, A.B. Kuzmenko, N.P. Armitage, N. Reyren, H. Hagemann, and I.I. Mazin, Phys. Rev. Lett. {\bf 100}, 226403 (2008);
R. Cowley, Phys. Rev. {\bf 134}, A981 (1964).

\bibitem{strain}
A. Janotti, D. Steiauf, and C.G. Van de Walle, Phys. Rev. B {\bf 84}, 201304 (2011);
Y.J. Chang, G. Khalsa, L. Moreschini, A.L. Walter, A. Bostwick, K. Horn, A.H. MacDonald, and E. Rotenberg, Phys. Rev. B {\bf 87}, 115212 (2013).

\bibitem{Allen} S.J. Allen, D.C. Tsui, and B. Vinter, Solid State Commun. {\bf 20}, 425 (1976).






\end{thebibliography}
\end{document}